\documentclass[11pt,a4paper,notoc]{article}
\pdfoutput=1
\usepackage{jcappub}
\usepackage{slashed}

\title{\begin{boldmath} Loop-induced dark matter direct detection signals from $\gamma$-ray lines \end{boldmath}}
\author{Mads T.~Frandsen,}
\author{Ulrich Haisch,}
\author{Felix Kahlhoefer,}
\author{\phantom{}\\ Philipp Mertsch}
\author{and Kai Schmidt-Hoberg} 
\affiliation{Rudolf Peierls Centre for Theoretical Physics,
  University of Oxford, \\ 1 Keble Road, Oxford OX1 3NP, United Kingdom}
\emailAdd{m.frandsen1@physics.ox.ac.uk}
\emailAdd{u.haisch1@physics.ox.ac.uk}
\emailAdd{felix.kahlhoefer@physics.ox.ac.uk}
\emailAdd{p.mertsch1@physics.ox.ac.uk}
\emailAdd{ksh@physics.ox.ac.uk}

\def \refeq#1{(\ref{#1})}

\arxivnumber{OUTP-12-12-P}

\abstract{ Improved limits as well as tentative claims for dark matter annihilation into $\gamma$-ray lines have been presented recently. We study the direct detection cross section induced from dark matter annihilation into two photons in a model-independent fashion, assuming no additional couplings between dark matter and nuclei. We find a striking non-standard recoil spectrum due to different destructively interfering contributions to the dark matter nucleus scattering cross section. While in the case of $s$-wave annihilation the current sensitivity of direct detection experiments is insufficient to compete with indirect detection searches, for $p$-wave annihilation the constraints from direct searches are comparable.  This will allow to test dark matter scenarios with $p$-wave annihilation that predict a large di-photon annihilation cross section in the next generation of experiments. 
}


\date{\today}

\notoc

\begin{document}

\maketitle

\section{Introduction}

Gravitational effects on astrophysical scales give convincing evidence for the presence of dark matter (DM) in our Universe, an observation which is strongly supported by measurements of the cosmic microwave background anisotropies \cite{Jarosik:2010iu}. While the existence of DM seems firmly established, very little is known about the properties of the DM particle(s). In order to identify their nature, several complementary search strategies are currently being employed, which fall into three classes: direct detection in shielded underground detectors, indirect detection with satellites, balloons and ground based telescopes looking for DM annihilation signals and DM production at colliders such as the LHC.

In general, indirect detection experiments looking for DM annihilations have to deal with astrophysical backgrounds, which make it difficult to unambiguously extract DM signals. One exception is annihilation into mono-energetic photons, which would provide a striking evidence for DM in our Galaxy. Consequently, $\gamma$-ray lines are among the most important search channels for the indirect detection of DM~\cite{Ackermann:2012qk}. Recently, there have been hints at a tentative $\gamma$-ray line at $E_{\gamma} \simeq 130 \, {\rm GeV}$ associated with a rather large annihilation cross section of  $\langle \sigma_{\chi} \hspace{0.25mm} v_{\rm rel} \rangle \simeq  1.3 \cdot 10^{-27} \, {\rm cm}^3 \, {\rm s}^{-1}$~\cite{Bringmann:2012vr, Weniger:2012tx} (assuming an Einasto profile, see subsequently also~\cite{Tempel:2012ey,Su:2012ft}). Such a $\gamma$-ray line has been predicted by various authors (see e.g.~\cite{Gustafsson:2007pc,Dudas:2009uq,Jackson:2009kg,Arina:2009uq}). This finding, although in slight tension with the upper limit $\langle \sigma_{\chi} \hspace{0.25mm} v_{\rm rel} \rangle \lesssim 1.0 \cdot 10^{-27} \, {\rm cm}^3 \, {\rm s}^{-1}$ set by the Fermi LAT Collaboration~\cite{Ackermann:2012qk}, has triggered a noticeable amount of theoretical studies~\cite{Chalons:2011ia,Chalons:2012hf,Ibarra:2012dw,Dudas:2012pb,Cline:2012nw,Choi:2012ap,Kyae:2012vi,Lee:2012bq,Rajaraman:2012db,Buckley:2012ws,Chu:2012qy,Das:2012ys,Kang:2012bq,Weiner:2012cb,Buchmuller:2012rc,Cohen:2012me,Cholis:2012fb}  most of which aim at explaining the possible $\gamma$-ray signal in terms of new physics. 

In the following we show that in view of the impressive sensitivity of current (and upcoming) direct detection experiments, relevant constraints on the nature of DM can arise from these searches  even if the interactions of DM with quarks and gluons are loop suppressed.  In particular,  we will explore the constraints on the annihilation cross section of DM into photons that arise from direct detection bounds. In order to keep our discussion as general as possible, we will work in an effective field theory (EFT) framework obtained by integrating out heavy degrees of freedom above a certain high-energy scale. Our model-independent analysis includes the  most relevant effective operators of dimension up to 7, containing bilinears constructed from scalar and fermion DM fields which lead to annihilations into $\gamma\gamma$.

We find that in the case of annihilation via $s$-wave, the constraining power of present direct detection searches is far below that of  indirect detection experiments. For $p$-wave annihilation, on the other hand, the constraints from direct and indirect searches are competitive. Given the expected improvement of direct detection experiments, DM models with $p$-wave annihilation and large di-photon cross section will hence be testable in the near future. This finding underscores the  possible complementary between direct and indirect searches in unraveling the precise nature of DM. 

Our work is organised as follows.  After listing the relevant effective operators that can give rise to a mono-energetic $\gamma$-ray line in Section~\ref{sec:operators}, we calculate in Section~\ref{sec:annihilation} the corresponding annihilation cross sections into a pair of photons.  Section~\ref{sec:third} is devoted to a comprehensive discussion of the loop-induced effective interactions relevant for DM direct detection. In this section we also estimate the associated nuclear matrix elements, assuming the dominance of a single operator. Applying recent bounds from DM direct detection experiments, we finally obtain in Section~\ref{sec:directdetection}  lower limits on the suppression scale of the effective operators, which in turn translate into model-independent upper bounds on the $\gamma \gamma$ annihilation cross section. A summary of our main results is presented in Section~\ref{sec:discussion}. A series of appendices contains useful details concerning technical aspects of our calculations.

\section{Operator analysis}
\label{sec:operators}

Throughout our analysis we will assume that both DM annihilation and direct searches can be described in terms of effective operators that are generated by integrating out heavy degrees of freedom. While such an EFT provides an excellent description of the low-energy processes involved in direct detection, the mass scale $M_\ast$ of new physics should be sufficiently high for the  framework to be applicable for DM annihilation. In particular, scenarios  with resonant $s$-channel annihilation are not covered by our analysis.

In our work we consider only effective operators that lead to the process $\chi \chi \rightarrow \gamma \gamma$, where $\chi$ denotes the DM particle. As pointed out in~\cite{Rajaraman:2012db}, dimension-4 operators correspond to milli-charged DM, which is strongly constrained experimentally and will hence not be discussed here. If DM is a Dirac fermion $D$, the leading contributions then arise from dimension-5 operators of electric or magnetic dipole type. Such operators give rise to longrange interactions between DM and nucleons at tree level and therefore to a sizeable direct detection cross section. The resulting constraints have been studied e.g.~in~\cite{Barger:2010gv,Banks:2010eh}, and it turns out that existing direct detection experiments constrain the annihilation cross section into $\gamma\gamma$ to many orders of magnitude below the thermal cross section. An observable $\gamma$-ray line can hence not arise from such operators and we will  therefore  not consider these interactions any further. The same line of reasoning applies to other higher-dimensional operators  that induce a tree-level coupling of DM to nucleons like the charge radius operator for complex scalar DM.  Consequently, we will restrict ourselves to operators that are bilinear in both the DM  and the photon fields.

For real scalar DM $R$, we obtain two different dimension-6 operators
\begin{align} \label{eq:operators1}
\mathcal{O}^R & = \mathcal{C}^R R^2 F^{\mu \nu} F_{\mu \nu} \,, & 
\mathcal{O}^R_\epsilon & = \mathcal{C}^R_\epsilon R^2 F^{\mu \nu} \tilde{F}_{\mu \nu} \,,
\end{align}
where  $F^{\mu \nu} = \partial^\mu A^\nu - \partial^\nu A^\mu$ denotes the usual electromagnetic field strength tensor, while $\tilde{F}^{\mu\nu}=\frac{1}{2}\epsilon^{\mu\nu\rho\sigma}F_{\rho\sigma}$  with $\epsilon^{0123} =1$ is its dual.  The corresponding operators for complex scalar DM $C$, called $\mathcal{O}^C$ and $\mathcal{O}^C_\epsilon$, are obtained from the above by replacing $R^2$ with $C^\dagger C$. Notice that  due to dimensional reasons the Wilson coefficients $\mathcal{C}^{R,C}_{(\epsilon)}$ appearing in~(\ref{eq:operators1}) all scale as $M_\ast^{-2}$.  

For fermionic DM, the relevant  operators are of dimension 7. In the case of Majorana fields $M$, we obtain two different sets of operators depending on whether the interactions entail the scalar $\bar{M} M$ or the pseudoscalar current  $\bar{M} \gamma^5 M$:
\begin{align} \label{eq:operators3}
\mathcal{O}^{M}& = \mathcal{C}^M \bar M M F^{\mu \nu} F_{\mu \nu} \,, &
\mathcal{O}^{M}_{\epsilon} & = \mathcal{C}^M_{\epsilon} \bar M M F^{\mu \nu} \tilde{F}_{\mu \nu} \,, \nonumber \\[-2.5mm] \\[-2.5mm]
\mathcal{O}^{M}_\text{p} & = \mathcal{C}^M_\text{p} \bar M \gamma^5 M F^{\mu \nu} F_{\mu \nu} \,, &
\mathcal{O}^{M}_{\text{p} \epsilon} & = \mathcal{C}^M_{\text{p}\epsilon} \bar M \gamma^5 M F^{\mu \nu} \tilde{F}_{\mu \nu} \,. \nonumber 
\end{align}
 For a Dirac fermion, the corresponding operators are again simply obtained by replacing $M$ with $D$. Additionally, if DM is a Dirac fermion, we can write down an operator involving the tensor current $\bar D \sigma^{\mu \nu} D$, where $\sigma^{\mu \nu} = \frac{i}{2}\left(\gamma^\mu\gamma^\nu-\gamma^\nu\gamma^\mu\right)$. It reads  
\begin{align}  \label{eq:operators4}
\mathcal{O}^{D}_\text{t} & = \mathcal{C}^D_\text{t} \bar D \sigma^{\mu \nu} D F_{\phantom{}\mu}^{\phantom{\nu} \rho} \tilde{F}_{\nu \rho} \, .
\end{align}
In contrast to the case of scalar DM, the Wilson coefficients entering the definitions (\ref{eq:operators3}) and (\ref{eq:operators4}) are proportional to $M_\ast^{-3}$. A summary of all the effective operators considered in this paper together with their main properties is given in Table~\ref{tab:overview}.

\begin{table}[tb]
\setlength{\tabcolsep}{5pt}
\renewcommand{\arraystretch}{1.6}
\center
\begin{tabular}{||c|c|c|c||} 
\hline \hline 
Operator & Scaling  &  Annihilation & Direct detection
\\
\hline
$\mathcal{O}^R$ & $M_\ast^{-2}$  & $\tfrac{8}{\pi} m_{R}^2 \left | \mathcal{C}^{R} (M_\ast) \right |^2$
 & $ \tfrac{4}{\pi} \mu_A^2 \left|f^R \right|^2 m_R^{-2} $
\\
\hline
$\mathcal{O}^R_\epsilon$ & $M_\ast^{-2}$  & $\tfrac{8}{\pi} m_{R}^2 \left | \mathcal{C}^{R}_\epsilon (M_\ast) \right |^2$ &  0
\\
\hline
$\mathcal{O}^M$ & $M_\ast^{-3}$  & $\tfrac{4}{\pi} v_\text{rel}^2 m_{M}^4 \left | \mathcal{C}^{M} (M_\ast) \right |^2$ & $  \tfrac{4}{\pi} \mu_A^2 \left|f^M \right|^2$
\\
\hline
$\mathcal{O}^M_{\epsilon}$ & $M_\ast^{-3}$  & $\tfrac{4}{\pi} v_\text{rel}^2 m_{M}^4 \left | \mathcal{C}^{M}_{\epsilon} (M_\ast) \right |^2$ & 0
\\
\hline
$\mathcal{O}^M_\text{p}$ & $M_\ast^{-3}$  & $\tfrac{16}{\pi} m_{M}^4 \left | \mathcal{C}^{M}_{\text{p}} (M_\ast) \right |^2$ & Suppressed
\\
\hline
$\mathcal{O}^M_{\text{p}\epsilon}$ & $M_\ast^{-3}$  & $\tfrac{16}{\pi} m_{M}^4 \left | \mathcal{C}^{M}_{\text{p}\epsilon} (M_\ast) \right |^2$ &  Suppressed
\\
\hline
$\mathcal{O}^D_\text{t}$ & $M_\ast^{-3}$  & $ \tfrac{1}{8 \pi} v_\text{rel}^2 m_{D}^4 \left | \mathcal{C}^{D}_{\text{t}} (M_\ast) \right |^2$ & Spin dependent
\\
\hline \hline
\end{tabular}
\caption{\label{tab:overview} Effective operators constructed out of DM  and field strength bilinears together with their main properties. See Section~\ref{sec:annihilation} for a discussion of the di-photon annihilation cross sections and Section~\ref{sec:directdetection} for details on the direct detection cross sections. The cross sections for complex scalar (Dirac fermion) DM  are smaller than the ones for real scalar (Majorana fermion) DM by a factor of~4.}
\vspace{-0.35cm}
\end{table}

\section{Annihilation cross sections}
\label{sec:annihilation}

In order to calculate the di-photon annihilation cross section of DM, we have to first compute the squared matrix elements for the process $\chi  \chi \to \gamma \gamma$ with $\chi = R, C, D, M$. For the set of operators introduced in the previous section, we find 
\begin{equation} \label{eq:annihilation}
\begin{split}
\left |{\cal M}^S_{(\epsilon)} \right |^2_{\gamma \gamma} & = \left | \mathcal{C}^{S}_{(\epsilon)} (M_\ast)  \right |^2  s^2 \cdot \begin{cases} 32 \,,  & S = R \,, \\
8 \,, & S= C \,,  \end{cases}  \\
\left |{\cal M}^F_{(\epsilon)} \right |^2_{\gamma \gamma}  & =  \left | \mathcal{C}^{F}_{ (\epsilon) } (M_\ast)  \right |^2  s^2 \left (s - 4 m_F^2 \right ) \cdot \begin{cases} 4 \,,  & F = D \,, \\
16 \,, & F= M \,,  \end{cases}  \\
 \left |{\cal M}^{F}_{\text{p} (\epsilon)} \right |^2_{\gamma\gamma} & =  \left |\mathcal{C}^{F}_{\text{p}  (\epsilon)} (M_\ast)  \right |^2 s^3 \cdot \begin{cases} 4 \,,  & F = D \,, \\
16 \,, & F= M \,,  \end{cases}  \\
  \Big |{\cal M}^{D}_\text{t} \Big |^2_{\gamma\gamma} & = \frac{1}{2} \, \Big |\mathcal{C}^{D}_\text{t} (M_\ast) \Big |^2 s^2 \left (s - 4 m_D^2 \right ) \,,
\end{split}
\end{equation}
where $s = 4 m_\chi^2 + m_\chi^2 v_{\rm rel}^2 + {\cal O} (v_{\rm rel}^4)$ denotes the center of mass energy.
 In the following we will assume $v_{\rm rel} \simeq 1.3 \cdot 10^{-3} \, c$. We discuss the uncertainties concerning $v_{\rm rel}$ in Appendix~\ref{app:vrel}.

In terms of the amplitudes (\ref{eq:annihilation}) the velocity-averaged annihilation cross section can be written as
\begin{equation} \label{eq:annihilationcrosssection}
\langle \sigma^\chi_i v_\text{rel} \rangle_{\gamma \gamma} = \frac{\left |\mathcal{M}^\chi_i \right |^2_{\gamma \gamma}}{64 \pi m_\chi^2} \,,
\end{equation}
where the label $i$ indicates the type of operator insertion and an additional factor of $1/2$ has been taken into account reflecting annihilation into two indistinguishable particles. The explicit results for the velocity-averaged annihilation cross section for each operator can be found in Table~\ref{tab:overview}. Note that Dirac and complex scalar DM are not their own anti-particles resulting in an additional factor of $1/2$ when calculating the $\gamma$-ray flux from the annihilation rate, assuming equal amounts of DM particles and anti-particles.

Although we are ultimately interested in deriving bounds on the allowed di-photon annihilation signal imposed by direct detection, it seems worthwhile to estimate the scale $M_\ast$ suppressing the higher-dimensional operators. This can be done by setting the DM annihilation cross sections into $\gamma \gamma$ equal to the benchmark value $\langle \sigma_{\chi} \hspace{0.25mm} v_{\rm rel} \rangle_{\gamma \gamma} \simeq 10^{-27} \, {\rm cm}^3 \, {\rm s}^{-1}$,   and then solving for the Wilson coefficients $\mathcal{C}^\chi_i$. Since the higher-dimensional operators we consider can only be generated from loops of new particles charged under $U(1)_{\rm em}$, one obtains $M_\ast$ of around $100 \, {\rm GeV}$ and $10 \, {\rm GeV}$ for $s$- and $p$-wave annihilation, respectively.  Note that this effective suppression scale does not directly correspond to the scale of new physics, because couplings as well as the multiplicities of new states have to be taken into account. The above numbers nevertheless suggest that scenarios leading to  an observable $\gamma$-ray signal are difficult to construct. In fact, a model that gives rise to $\langle \sigma_{\chi} \hspace{0.25mm} v_{\rm rel} \rangle_{\gamma \gamma}  \sim 10^{-27} \, {\rm cm}^3 \, {\rm s}^{-1}$ should either contain a high multiplicity of states running in the loops that generate the effective operators or should feature resonant ($s$-channel) production in which case the mass scale $M_\ast$ gets replaced by the width $\Gamma_\ast$ of the mediator that is exchanged in the annihilation process. We will not discuss possible ultraviolet (UV) completions.

\section{Loop-induced effective interactions in direct detection}
\label{sec:third}

The  dimension 6 and 7 operators ${\cal O}_i^\chi$ introduced in Section~\ref{sec:operators} lead to interactions between DM particles and nuclei which can be probed in direct detection experiments. There are in essence two kinds of effects which contribute to the DM-nucleus scattering cross section. The first  one results from loop-induced couplings of DM to individual quarks and gluons, while the second one corresponds to Rayleigh scattering which stems from coherent interactions of the two photons with the entire nucleus~\cite{Weiner:2012cb}. As we explain below, the different contributions involve several widely separated energy scales, namely the high-energy scale $M_\ast$,  characterising the onset of non-standard dynamics, the heavy-quark thresholds $m_Q$, the light-quark masses $m_q$, the inverse  nuclear coherence length $Q_0$, and finally the momentum exchange $q$, involved in the low-energy scattering. In order to separate short- from long-distance physics, it will be useful to  again employ an EFT. 

In the following we will present the basic steps of this EFT calculation, quoting explicit results for the operator ${\cal O}^M$ only. Further details on the computation of the individual contributions itself are relegated to Appendix~\ref{app:running}. The  generalisation to the case of the other operators is straightforward and will be presented at the end of the section. For the further discussion we also make the assumption that the scale of new physics $M_\ast$ is above the top-quark threshold $m_t$.

\subsection{Operator mixing and threshold corrections}
\label{sec:mixing}

\begin{figure}
\begin{center}
\includegraphics[width=13cm]{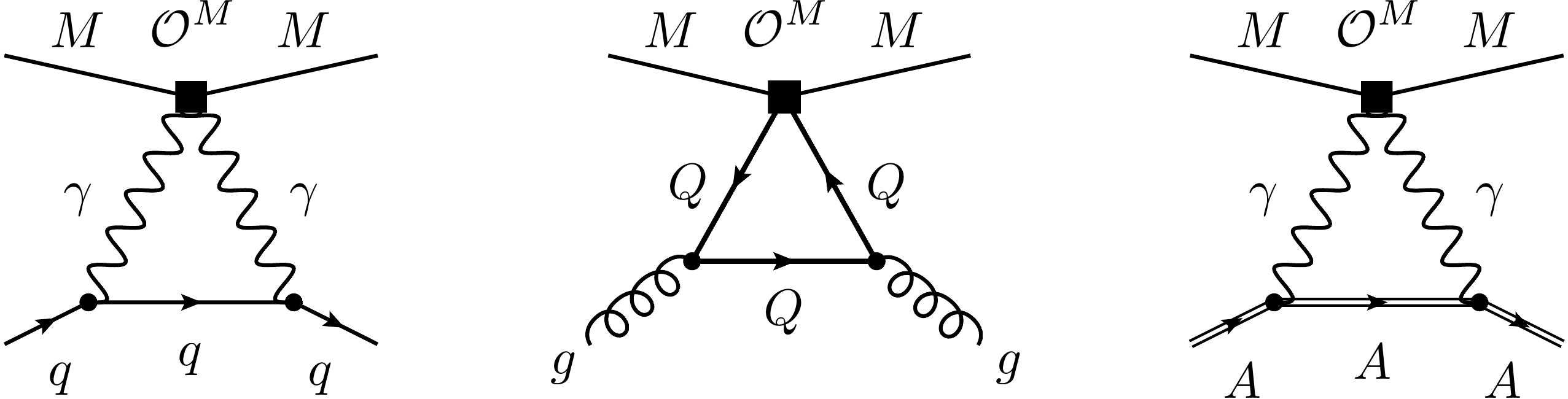}
\end{center}
\vspace{-2mm}
\caption{\label{fig:diagrams} One-loop Feynman graphs showing the contributions to the DM-nucleus cross section in the case of ${\cal O}^M$. Mixing diagram generating  ${\cal O}_q^M$ (left), matching contribution giving rise to ${\cal O}_G^M$ (middle), and matrix element describing the low-energy two-photon scattering of DM on the nucleus (right). See text for further details.}
\end{figure}

The first important observation is that ${\cal O}^M$ mixes under QED into the operator 
\begin{equation} \label{eq:Qq}
{\cal O}^M_q = {\cal C}_q^M m_q \bar M M \bar q q \,, 
\end{equation}
where $m_q$ denotes the mass of the SM quark $q$, which can have any flavor $q = u,d, s, c, b, t$. This mixing is determined by   the UV pole of the Feynman diagram shown on the left-hand side in Figure~\ref{fig:diagrams}. To leading logarithmic accuracy, one finds the following relation between the Wilson coefficient of the operator of ${\cal O}_q^M$ and that of ${\cal O}^M$, 
\begin{equation} \label{eq:CqM}
\mathcal{C}_q^M (\mu) \simeq -3  e_q^2\, \frac{\alpha}{\pi}  \, \ln \left (\frac{M_\ast^2}{\mu^2}  \right ) \, \mathcal{C}^M  (M_\ast) \,,
\end{equation}
where $e_q$ is the electric charge of the quark $q$ and  $m_t < \mu < M_\ast$. Notice that we have assumed that the Wilson coefficient of ${\cal O}_q^M$ vanishes at $M_\ast$.  

We now evolve the Wilson coefficient ${\cal C}_q^M$ from $M_\ast$ down to $m_t$, where we integrate out the top quark. Removing the heavy quark as an active degree of freedom gives rise to a finite threshold correction to the Wilson coefficient of the operator
\begin{equation} \label{eq:CGM}
{\cal O}^M_G = {\cal C}_G^M  \bar M M G^{a, \mu \nu} G^a_{\mu \nu} \,,
\end{equation}
where $G^{a, \mu \nu}$ denotes the field strength tensor of QCD. The relevant leading-order (LO) diagram is shown in the middle of Figure~\ref{fig:diagrams}. The corresponding matching is captured by the simple replacement~\cite{Shifman:1978zn}
\begin{equation} \label{eq:SVZ}
m_t  \bar M M \bar t t \, \mathcal{C}_t^M (m_t) \to   \bar M M G^{a, \mu \nu} G^a_{\mu \nu} \, {\cal C}_G^M (m_t) \,,
\end{equation}
with  ${\cal C}_G^M$ given at next-to-leading order (NLO)  by 
\begin{equation} \label{eq:CGmt}
\mathcal{C}_{G}^M (m_t)  =  -\frac{\alpha_s (m_t)}{12 \pi}\, \big (  1 + \delta_t  \big ) \, \mathcal{C}_t^M  (m_t) \, ,
\end{equation}
where $\delta_t =  11 \alpha_s (m_t)/(4 \pi)$~\cite{Inami:1982xt}. Although $\delta_t$ is  formally of higher order, we will include such finite two-loop contributions in our analysis, because they are numerically non-negligible. Notice that once the top quark has been removed, the Wilson coefficient $ \mathcal{C}_t^M$ and the corresponding logarithm is frozen at the threshold $m_t$ in the EFT. 

After the top quark has been integrated out, we then have to consider the mixing of the set of three operators ${\cal O}^M$, ${\cal O}_{q}^M$ and ${\cal O}_{G}^M$. Like  ${\cal O}^M$ the operator  ${\cal O}_{G}^M$ mixes into ${\cal O}_{q}^M$.  The relevant diagram is the QCD counterpart of the one displayed on the left in Figure~\ref{fig:diagrams} with the photons replaced by gluons. As shown in Appendix~\ref{app:running}, the associated corrections are subleading and we will neglect them in what follows. The operator ${\cal O}_{G}^M$ itself evolves like the QCD coupling constant, so that for scales  $m_b < \mu < m_t$ its Wilson coefficient takes the form 
\begin{equation} \label{eq:CG}
\mathcal{C}_G^M (\mu)  \simeq \frac{\alpha}{\pi}  \,  \frac{\alpha_s (\mu)}{ \pi} \,  \frac{e_t^2}{4} \,  \big (  1 + \delta_t  \big ) \, \ln \left ( \frac{M_\ast^2}{m_t^2} \right )  \mathcal{C}^M (M_\ast) \,.
\end{equation}

At the scales $m_b$ and $m_c$, the bottom and charm quarks are integrated out, which in full analogy to  \refeq{eq:CGmt} results in finite matching corrections to ${\cal C}_G^M$. Including all heavy-quark threshold effects as well as the renormalisation group (RG) evolution of the individual Wilson coefficients, we find for $\mu < m_c$  the following expressions
\begin{equation} \label{eq:Clow}
\begin{split}
\mathcal{C}^M (\mu) & \simeq \mathcal{C}^M (M_\ast)  \,, \\
\mathcal{C}_q^M (\mu) & \simeq -3  e_q^2\, \frac{\alpha}{\pi}  \, \ln \left (\frac{M_\ast^2}{\mu^2}  \right ) \, \mathcal{C}^M  (M_\ast) \,, \\
\mathcal{C}_G^M (\mu)  & \simeq \frac{\alpha}{\pi}  \,  \frac{\alpha_s (\mu)}{ \pi} \,  \sum_{Q} \frac{e_Q^2}{4} \, \big (  1 + \delta_Q \big ) \, \ln \left ( \frac{M_\ast^2}{m_Q^2} \right )  \mathcal{C}^M (M_\ast) \,,
\end{split}
\end{equation}
with $q$ any of the light quarks $u,d,s$, while the sum over $Q$ includes the heavy quarks $c,b,t$.

\subsection{Nuclear matrix elements}
\label{sec:nuclear}

Having determined the values of the Wilson coefficients at the hadronic scale $\mu = {\cal O}(1 \, {\rm GeV})$, we still have to evaluate the matrix elements of the operators ${\cal O}^M$, ${\cal O}_{q}^M$ and ${\cal O}_{G}^M$ between nucleus states $A$. We begin our discussion with the matrix element $\langle A | {\cal O}^M | A \rangle$. The important observation in this context~\cite{Weiner:2012cb} is that in general the two virtual photons do not scatter off a single nucleon, but interact with the entire nucleus (see also~\cite{Pospelov:2000bq} for a related discussion). This is illustrated by the Feynman diagram on the right in Figure~\ref{fig:diagrams}. The strength of this interaction is to first order a measure of the total electric charge $Ze$ of the nucleus, while effects related to its substructure  (such as spin or magnetic moment) are suppressed by the nucleus mass. The matrix element of $ {\cal O}^M$ can hence be computed by treating the nucleus as a slowly moving charge source, applying well-known techniques of heavy-quark effective theory. The basic steps of the actual calculation can be found in Appendix~A of \cite{Weiner:2012cb} and will not be repeated here. To leading power in the nucleus mass and zeroth order in the velocity expansion, we find\footnote{In \refeq{eq:neal}, \refeq{eq:lightmatrix}, and \refeq{eq:heavymatrix} the spinors associated to the DM bilinears are suppressed.}
\begin{equation} \label{eq:neal} 
\langle A | {\cal O}^M | A \rangle \simeq 2 \hspace{0.25mm} \sqrt{\frac{2}{\pi}} \, \alpha Z^2 Q_0 \hspace{0.25mm} F_\text{Ray}(\bar q) \hspace{0.5mm}  {\cal C}^M (\mu) \,,
\end{equation}
 where we have introduced the form factor\footnote{Our results \refeq{eq:neal} and~\eqref{eq:Fbarq} differ slightly from the ones reported in~\cite{Weiner:2012cb}. The difference can be traced back to their equation~(A-11), where a factor of 2 is missing in the exponent after substituting~(A-9) into~(A-7). We also note that the sine function in  (A-17) should be a hyperbolic  sine as in our formula \refeq{eq:Fbarq}.}
\begin{equation} \label{eq:Fbarq}
\begin{split}
F_\text{Ray}(\bar q) & = -\sqrt{\frac{2}{\pi}} \int_0^1 \! d x \int_0^\infty d l \; \frac{l^2}{\left(l^2 + \bar q^2 (1-x)x \right)^2} \,  \exp \left[ - 2 \left ( l^2 - \bar q^2 \left ( (1-x) x - \tfrac{1}{2} \right ) \right ) \right] \\[2mm] & \phantom{xx} \times \left[ {\rm cosh} \left(2 \hspace{0.25mm} l \bar q \left (1-2x \right )\right) - \frac{ l^2 - \bar q^2 (1-x)x +\tfrac{1}{2}}{ l \bar q\left  (1-2x \right )} \; {\rm sinh} \left(2 \hspace{0.25mm} l \bar q \left (1-2x \right ) \right)\right] \, ,
\end{split}
\end{equation}
 normalised such that $F_{\text{Ray}} (0) = 1$. Here  $\bar q =  |\boldsymbol{q}|/Q_0 = q/Q_0$ denotes the ratio of the three-momentum transfer  $\boldsymbol{q}$ and the nuclear coherence scale  $Q_0 \simeq 0.48 \hspace{0.5mm} (0.3 + 0.89 A^{1/3})^{-1} \, {\rm GeV}$. For xenon with $A \simeq  131.3$ one finds numerically  $Q_0 \simeq  0.1 \, {\rm GeV}$. Notice that the coherent contribution~\refeq{eq:neal} scales like $Z^2$ and is therefore more important for heavy nuclei.

We now turn to the calculation of the matrix elements $\langle A | {\cal O}_q^M | A \rangle$. Following the standard procedure in the DM literature, we evaluate these contributions at  tree level. Neglecting small differences between the nuclear matrix elements of protons and neutrons, we obtain  
\begin{equation} \label{eq:lightmatrix} 
\langle A | {\cal O}_{q}^M | A \rangle  \simeq 2 \hspace{0.25mm} m_N  A \hspace{0.25mm} f_{Tq}^N \hspace{0.5mm} F_\text{Helm}(\bar q) \hspace{0.5mm}   {\cal C}_q^M (\mu)  \,,
\end{equation} 
where $m_N \simeq 0.939 \, {\rm GeV}$ is the average nucleon mass and $A$ is the mass number of the nucleus, and the scalar form factor $f^N_{Tq}$ is defined via $m_N f^N_{Tq} = \langle N|m_q \bar{q}q|N\rangle$.  $F_\text{Helm}$ is the Helm form factor
\begin{equation} 
F_\text{Helm}(\bar q) = \frac{3 j_1(\bar q Q_0 R_A)}{\bar q Q_0 R_A}  \exp \left (- \tfrac{1}{2}  \bar q^2  Q_0^2  t^2 \right ) \,,
\end{equation}
 which depends on the effective nuclear radius $ R_A \simeq \sqrt{(6.1\, {\rm GeV}^{-1} \; A^{1/3})^2 - 5 t^2}$ and the nuclear skin thickness $t \simeq 4.6 \, {\rm GeV}^{-1}$, while $j_1$ denotes the spherical Bessel function of the first kind. Notice that a possible momentum dependence of the scalar form factors $f_{Tq}^N$  is not included in~\refeq{eq:lightmatrix}. Such effects together with other corrections arising at NLO have recently been computed in~\cite{Cirigliano:2012pq} employing the formalism of chiral perturbation theory (ChPT). While these corrections can be significant if there are large cancellations between the contributions from protons and neutrons, in our case their inclusion would have a minor impact only. The approximate result in equation~\refeq{eq:lightmatrix} is hence sufficient for our purposes.

The last missing ingredient in the calculation of the DM-nucleus cross section is  the nuclear matrix element of the operator ${\cal O}_{G}^M $. In contrast to~\refeq{eq:lightmatrix}, the first chiral corrections to  $\langle A | {\cal O}_{G}^M | A \rangle$ arise at the NNNLO level~\cite{Cirigliano:2012pq}. It is hence an excellent approximation to use the classic tree-level result of~\cite{Shifman:1978zn}. In our notation, one obtains 
\begin{equation} \label{eq:heavymatrix} 
\langle A | {\cal O}_{G}^M | A \rangle  \simeq - 2 \, \frac{8\pi}{9\alpha_s(\mu)} \, m_N  A \hspace{0.25mm} f_{TG}^N \hspace{0.5mm}  F_\text{Helm}(\bar q) \hspace{0.5mm} {\cal C}_G^M (\mu)  \,,
\end{equation}
where again the scalar form factors for protons and neutrons have been set equal. Following common practice, we have used the expression $f_{TG}^N = 1 - \sum_q f_{Tq}^N$ for the gluon form factor.  

The nuclear matrix elements for the operator  $\mathcal{O}^{R}$ are identical to the results given above, while those for ${\cal O}^{C,D}$ are smaller by a factor of 2 than \refeq{eq:neal},  \refeq{eq:lightmatrix} and \refeq{eq:heavymatrix}.
The contributions from the operators $\mathcal{O}^{\chi}_\epsilon$, on the other hand, all vanish due to the anti-symmetry of the Levi-Civita tensor in ${\tilde F}_{\mu \nu}$. Finally,  the pseudo-scalar operators $\mathcal{O}^{M, D}_{\text{p}(\epsilon)}$ and the tensor operator $\mathcal{O}^D_\text{t}$ generate only momentum-suppressed and spin-dependent interactions between quarks and DM at the one-loop level, respectively, which are irrelevant for direct detection.

\section{Direct detection cross sections}
\label{sec:directdetection}

We have seen in the previous section that only the dimension-6 operators ${\cal O}^{R,C}$ and the dimension-7 operators ${\cal O}^{M,D}$ lead to unsuppressed spin-independent (SI) interactions between DM particles and nuclei.  For these cases the scattering cross section on  the entire nucleus can be written as 
\begin{equation} \label{eq:sigmaSI}
\sigma^\chi_{\rm SI} = n_\chi^2 \,  \frac{ \mu^2_A}{\pi} \,\left |f^\chi \right|^2  \cdot \begin{cases} m_\chi^{-2} \,, & \chi = R,C \,, \\ 1 \,, & \chi = M,D \,,  
\end{cases}
\end{equation}
where $\mu_A = m_A m_\chi (m_A + m_\chi)^{-1}$ is the reduced mass of the DM-nucleus system and  $n_\chi=2$ for $\chi=R,M$, while $n_\chi=1$ for $\chi=C,D$. The quantity $f^\chi$ takes the general form 
\begin{equation} \label{eq:fchi}
f^\chi = \frac{1}{n_\chi}\, \Big ( \langle A | \mathcal{O}^\chi | A \rangle  + \sum_{q} \langle A | \mathcal{O}_q^\chi | A \rangle +  \langle A | \mathcal{O}_G^\chi | A \rangle \Big) \, ,
\end{equation}
and includes the three contributions evaluated in the previous section. 

\begin{figure}
\begin{center}
\includegraphics[width=0.46\columnwidth]{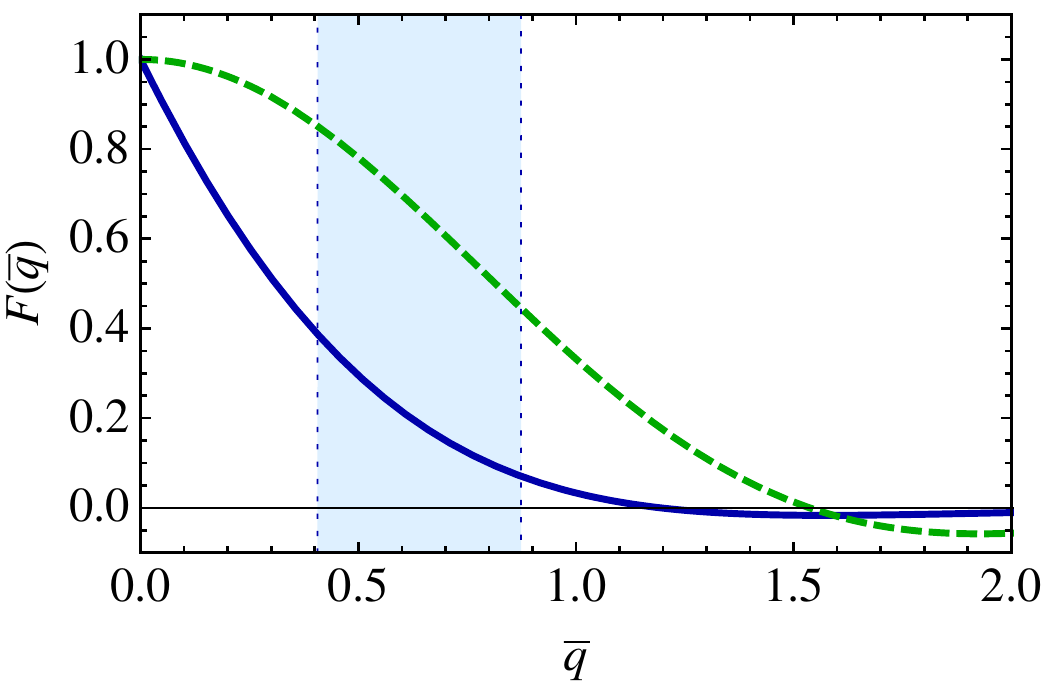}
\quad 
\includegraphics[width=0.49\columnwidth]{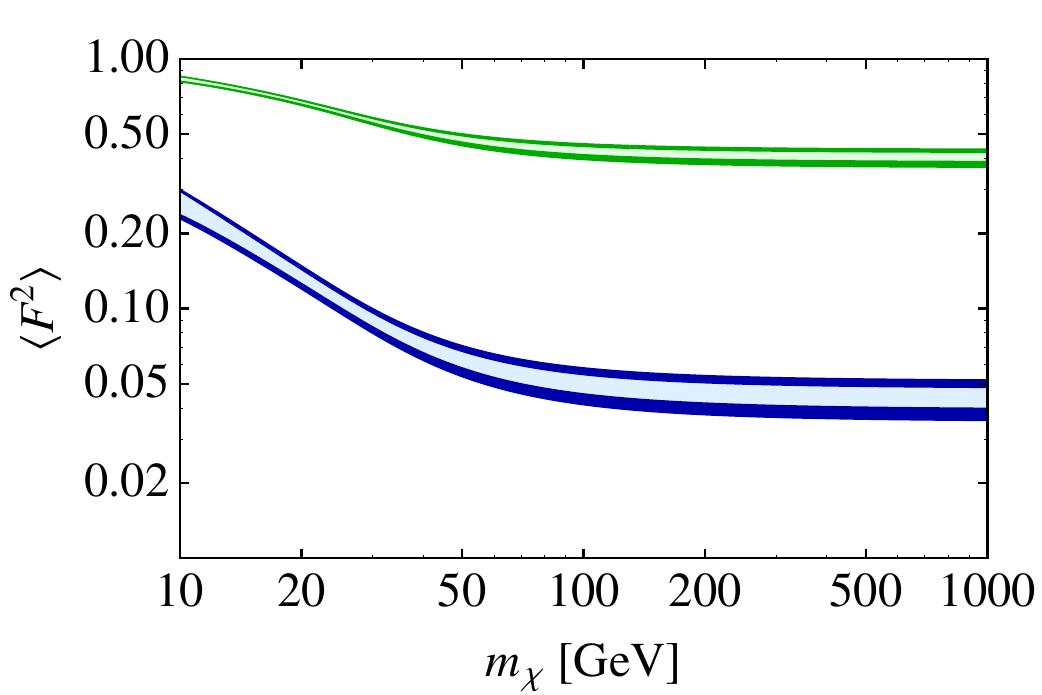}
\caption{\label{fig:formfactors}  Left panel: Comparison of the Rayleigh form factor (solid blue line) and the Helm form factor (dashed green line) as a function of $\bar q$. The Rayleigh form factor shows a much stronger suppression towards large momentum transfer. The blue shaded region displays the typical values of $\bar q$ relevant for the XENON100 detector. Right panel: Recoil energy averaged form factors $\langle F^2 \rangle$ or XENON100 as a function of the DM mass. The blue (green) band represents the result for the Rayleigh (Helm) form factor.
See text for further details.}
\end{center}
\end{figure}

Combining the relations \refeq{eq:Clow} to \refeq{eq:heavymatrix} the following explicit expression for the  effective coupling $f^\chi$ is found 
\begin{equation} \label{eq:fN} 
\begin{split}  
f^\chi & = \frac{\alpha}{\pi} \left [   \sqrt{2\pi} \, Z^2\, Q_0 \hspace{0.25mm} F_\text{Ray}(\bar q) - 3 m_N A \left ( \frac{4}{9} \, f_{Tu}^N  + \frac{1}{9} \left (  f_{Td}^N  +  f_{Ts}^N \right ) \right ) L_N F_\text{Helm}(\bar q) \right. \\[1mm] & \left. \phantom{xxxxi} - \frac{2}{9} \, m_N A \hspace{0.25mm}  f_{TG}^N \left ( \frac{4}{9} \left ( \Delta_c L_c + \Delta_t L_t \right) + \frac{1}{9} \, \Delta_b L_b \right )F_\text{Helm}(\bar q) \right ] \mathcal{C}^\chi(M_\ast) \,,
\end{split}
\end{equation}
where  $L_P = \ln \left (M_\ast^2/m_P^{2} \right )$ with $P=N,c,b,t$ and $\Delta_Q = 1+ \delta_Q$. Following \cite{Weiner:2012cb},  we will use $\alpha \simeq 1/137$ in our numerical analysis. The light-quark matrix elements $f_{Tq}^N$ can either be determined phenomenologically from baryon masses and meson-baryon scattering data or  computed within lattice QCD (see~\cite{Freytsis:2010ne} for a concise review). We adopt the values
\begin{equation} \label{eq:fs}
 f_{Tu}^N \simeq 0.021 \,, \qquad f_{Td}^N \simeq 0.038\,, \qquad  f_{Ts}^N \simeq 0.013\,.
\end{equation}
The numbers for $f_{Tu}^N$ and $f_{Td}^N$ have been obtained by averaging the values for the proton and neutron scalar form factors as given in~\cite{Freytsis:2010ne}, whereas $f_{Ts}^N$ has been taken from the recent lattice QCD study~\cite{Takeda:2010id}. Compared to older values of typically $f_{Ts}^N \simeq 0.14$~\cite{Freytsis:2010ne}  the number given in~\refeq{eq:fs} is notably smaller. As  discussed below, the resulting uncertainty associated to this choice is small. Notice also that  in~\refeq{eq:fN} we have identified the renormalisation scale $\mu$ with the scale at which the matrix elements of the operators of Section~\ref{sec:operators} are matched onto ChPT or another low-energy effective Lagrangian describing the  DM-nucleon interactions. For  definiteness we have used the nucleon mass $m_N$ in this matching.  In our EFT calculation we employ the $\overline{\rm MS}$ scheme and correspondingly we take  $m_c \simeq 1.3 \, {\rm GeV}$,  $m_b \simeq 4.2 \, {\rm GeV}$  and $m_t \simeq 165 \, {\rm GeV}$ for the heavy-quark masses.  The associated threshold effects are $\Delta_c \simeq 1.35$, $\Delta_b \simeq 1.20$ and $\Delta_t \simeq 1.10$ for $\alpha_s (m_c) \simeq 0.399$,  $\alpha_s (m_b) \simeq 0.226$ and  $\alpha_s (m_t) \simeq 0.109$.

As an example, we calculate $f^\chi$ for interactions between the DM particles and xenon atoms. Using the parameters specified above, we find
\begin{equation} \label{eq:fchiexample} 
f^\chi \simeq \left\{1.69\, F_\text{Ray}(\bar q) - \left[0.57 + 0.09 \, \ln \left(\frac{M_\ast^2}{(200\,{\rm GeV})^2}\right)\right] F_\text{Helm}(\bar q) \right\} \, \mathcal{C}^\chi(M_\ast) \, {\rm GeV}\,.
\end{equation}
We add that employing $f_{Ts} \simeq 0.14$ instead of the value quoted in \refeq{eq:fs} would change the numerical coefficient $0.57$ into $0.64$, implying that the theoretical uncertainty in $f^\chi$ related to the strange-quark scalar form factor is about $10\%$. The errors on $f_{Tu}^N$ and $f_{Td}^N$ have even less impact on the obtained results.

Our result for $f^\chi$ has some features worth noting. We first observe from \refeq{eq:fN} and \refeq{eq:fchiexample} that there is a relative sign between the coherent contribution induced directly  by $\mathcal{O}^\chi$ and the corrections from $\mathcal{O}^\chi_q$ and $\mathcal{O}^\chi_G$ arising due to operator mixing and threshold corrections. The two types of effects hence interfere destructively. In fact, the extent of the interference depends sensitively on the value of the coherent form factor $F_\text{Ray}(\bar q)$ compared to the standard form factor $F_\text{Helm}(\bar q)$. The $\bar q$-dependence of both form factors is shown in Figure~\ref{fig:formfactors}. From the curves it is evident that  the coherent form factor is much more strongly suppressed for finite three-momentum transfer  than the Helm form factor. 

\begin{figure}
\begin{center}
\includegraphics[width=0.45\columnwidth,clip,trim=0 -2 0 0]{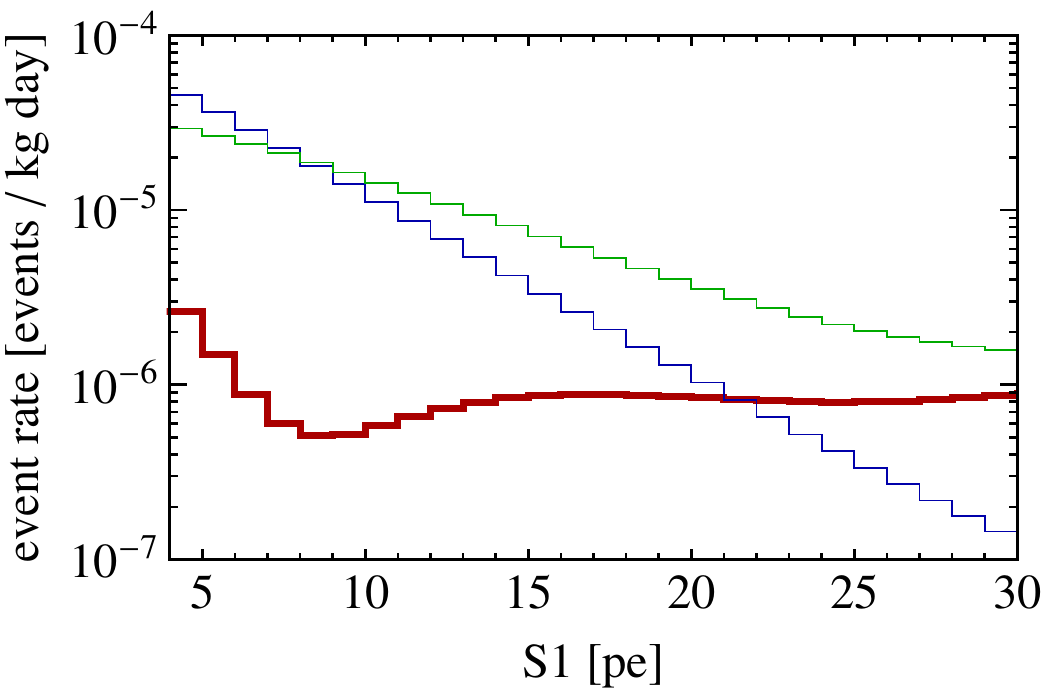}
\quad 
\includegraphics[width=0.46\columnwidth]{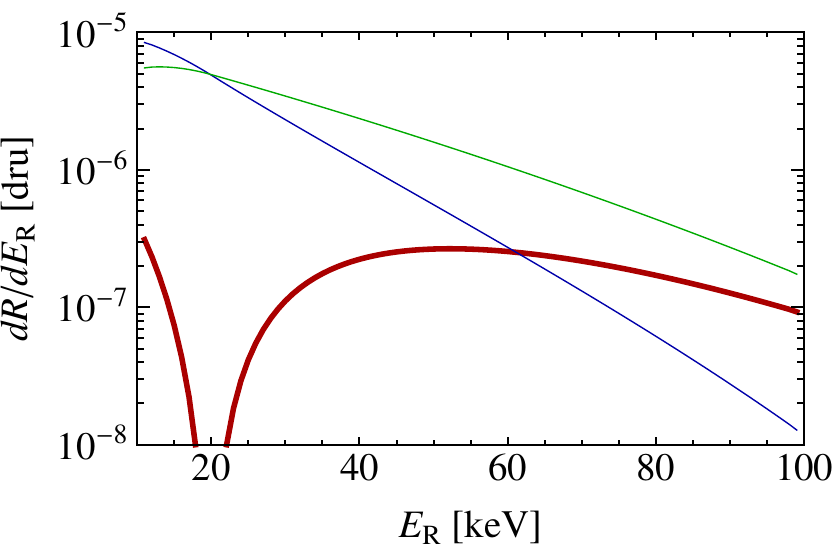}
\caption{
\label{fig:xenon1t}
Left panel: Expected event rate in XENON100 assuming $\mathcal{C}^\chi = 10^{-6}\, \rm{GeV}^{-3}$. The blue (green) line indicated the Rayleigh (mixing) contribution, while the red line illustrates the actual event rate after interference of the two contributions. Contrary to most DM models, the event rate does not decrease towards larger recoil energies, giving a distinctive signature of this model. Right panel: Differential event rate for CDMS-II, measured in ${\rm dru} = {\rm events}/({\rm kg} \, {\rm day} \, {\rm  keV})$. The displayed results again correspond to the choice $\mathcal{C}^\chi = 10^{-6}\, \rm{GeV}^{-3}$.}
\end{center}
\end{figure}

Clearly, a realistic direct detection experiment will not be able to probe the scattering cross section at zero-momentum transfer. Instead, it will only be sensitive to nuclear recoils with energy  $E_\text{R} =  q^2/(2 m_A)$ larger than the energy threshold $E_\text{th}$ of the detector. In order to make a realistic estimate of the relative importance of the two contributions we need to compare the two form factors at finite momentum transfer. In other words, we want to calculate the average value of the form factors, $\langle F^2 \rangle$, in a given direct detection experiment. This goal can be achieved by calculating the recoil energy average of the form factors for different values of $m_\chi$ as outlined in Appendix~\ref{app:xenon}.

The averaged Rayleigh and Helm form factors for XENON100~\cite{Aprile:2012} as a function of $m_\chi$ are displayed in the right panel of  Figure~\ref{fig:formfactors}. We see that the averaged Rayleigh form factor is significantly smaller than its Helm form factor counterpart.  The ratio of the two decreases from  $\langle F^2_\text{Ray} \rangle / \langle F^2_\text{Helm} \rangle \approx 0.3$ for $m_\chi \ll m_A$ to $\langle F^2_\text{Ray} \rangle / \langle F^2_\text{Helm} \rangle \approx 0.1$ for $m_\chi \gtrsim m_A$. To estimate the uncertainties of the averaged form factors, we have studied their dependence on the relative scintillation efficiency ${\cal L}_{\rm eff}$ in xenon and on the DM velocity distribution $f(v)$. We find that the error related to ${\cal{L}}_{\rm eff}$ is the dominant individual source of uncertainty in $\langle F^2 \rangle$ and amounts to roughly $10\%$ for large values of $m_\chi$. The associated $1\sigma$ ($2 \sigma$) error bands are displayed in  light (dark) colours in the right panel of Figure~\ref{fig:formfactors}.

\begin{figure}
\begin{center}
\includegraphics[width=0.45\columnwidth]{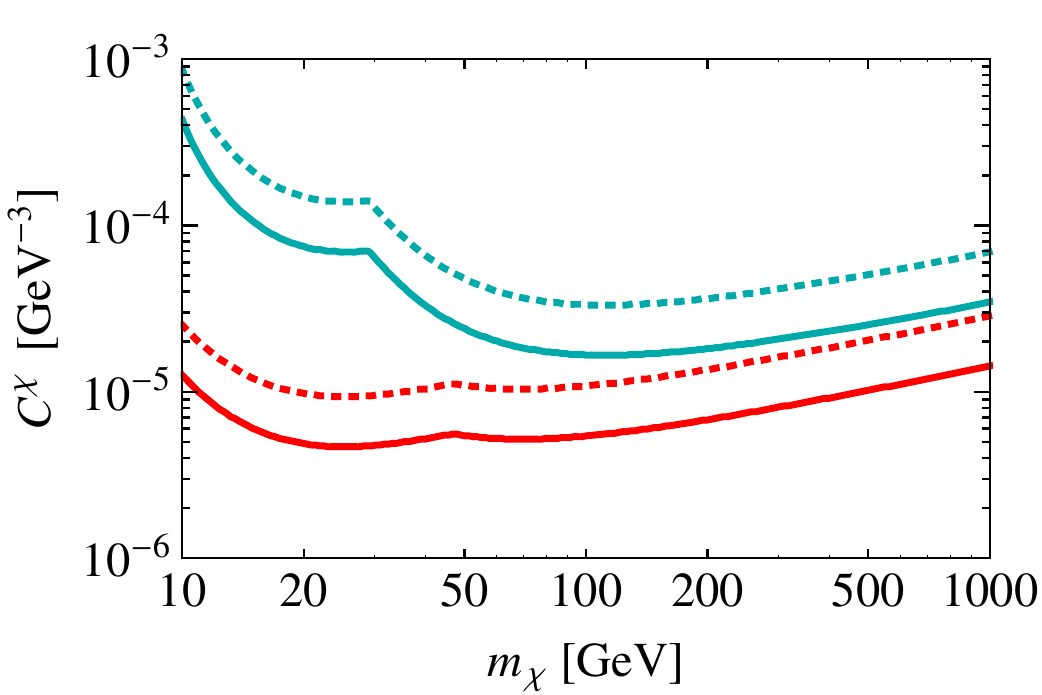}
\quad 
\includegraphics[width=0.45\columnwidth]{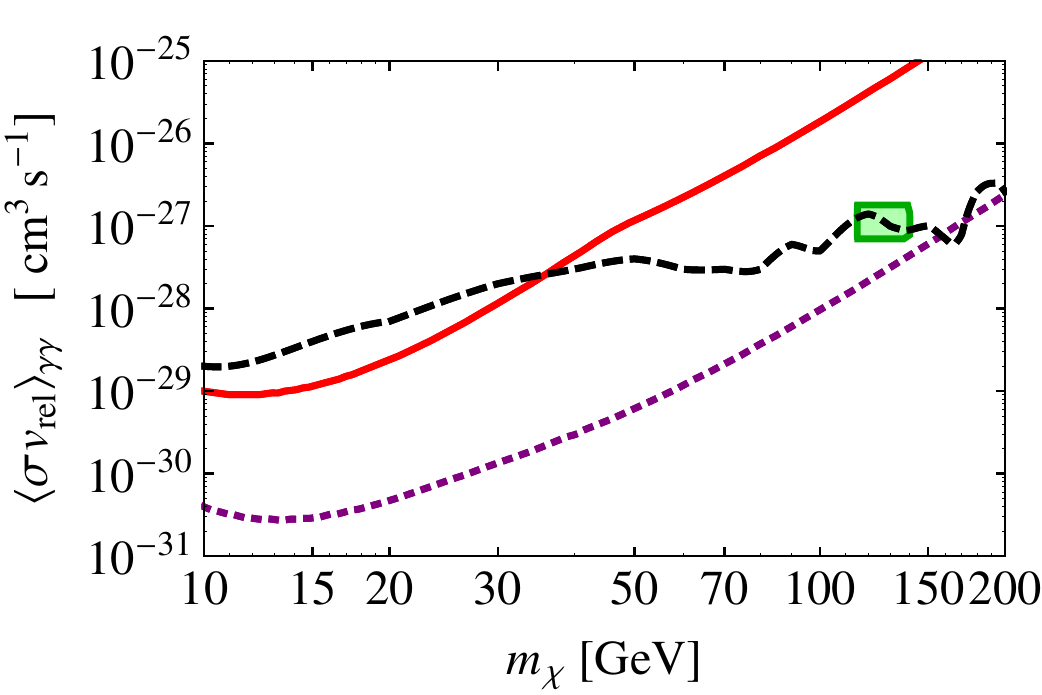}
\caption{
\label{fig:bounds}
Left panel: Limits on $C^\chi$ as a function of $m_\chi$ from XENON100 (red) and CDMS-II (blue) for $\chi = M$ (solid) and $\chi = D$ (dotted). The same limits apply to $C^\chi / m_\chi$ for $\chi = R$ (solid) and $\chi = C$ (dotted). Right panel: Bound on  $\langle \sigma v_{\rm rel} \rangle_{\gamma \gamma}$  as a function of $m_\chi$ from XENON100 (red solid curve) compared to the current bound from  Fermi LAT  (black dashed curve) for Majorana DM. The projected sensitivity of XENON1T is indicated by the purple dotted curve. The green box shows the parameters for the claimed $\gamma$-ray signal from~\cite{Weniger:2012tx}. For Dirac DM, the bounds from XENON100 and XENON1T would be stronger by a factor of 2. All shown curves have been obtained by setting the new-physics scale $M_\ast$ entering the effective coupling $f^\chi$ equal to $200 \, {\rm GeV}$.}
\end{center}
\end{figure}

Because of the strong suppression of the Rayleigh form factor for finite momentum transfer, the two terms in \eqref{eq:fchiexample} give comparable contributions to the differential event rate in a realistic detector.  Typically, the contribution proportional to the Rayleigh form factor is larger close to the threshold, while the contribution proportional to the Helm form factor dominates at large recoil energies. Consequently, there will be large interference effects leading to a distinct recoil spectrum. This striking feature is illustrated in Figure~\ref{fig:xenon1t} for both XENON100~\cite{Aprile:2012} (left panel) and CDMS-II~\cite{Ahmed:2009zw} (right panel). Note that the effect of the interference is much more pronounced in CDMS-II because of the much better energy resolution. Contrary to most DM models, we expect almost no events near the energy threshold, whereas the bulk of the signal is expected at relatively large momentum transfer.

In spite of the large interference, we still obtain relevant constraints on the Wilson coefficients $\mathcal{C}^\chi$ both from XENON100 and CDMS-II using the maximum gap method~\cite{Yellin}. The resulting bounds are shown in the left panel of Figure~\ref{fig:bounds} and range from around $10^{-3}  \, {\rm GeV}^{-3}$ to almost $10^{-6} \, {\rm GeV}^{-3}$. These constraints can be translated into bounds on the $\gamma \gamma$ annihilation cross section. In order to get a feeling for the resulting limits, we define $\sigma^\chi_0 = \sigma^\chi_\text{SI} (q^2 = 0)$ and $f_0 = f^\chi(q^2 = 0) / \mathcal{C}^\chi$. Combining \refeq{eq:annihilationcrosssection} with~\refeq{eq:sigmaSI}, we then obtain 
\begin{equation} \label{eq:sigmaSIbounds} 
\langle \sigma^S v_{\rm rel} \rangle_{\gamma \gamma} \lesssim n_S\, \frac{m_S^4}{\mu_A^2} \frac{\sigma^S_0}{f_0^2} \,, \qquad 
 \langle \sigma^F v_{\rm rel} \rangle_{\gamma \gamma} \lesssim \frac{n_F}{2} \, \frac{m_F^4}{\mu_A^2}  \frac{\sigma^F_0}{f_0^2} \, v_{\rm rel}^2\,, 
 \end{equation}
where $S=R,C$ and $F=M,D$. 

From the above relations it is readily seen that for scalar DM, corresponding to $s$-wave annihilation, the current sensitivity of direct detection searches is insufficient (by six to eight orders of magnitude) to compete with indirect detection experiments.
In contrast, for fermionic DM which leads to $p$-wave annihilation, the present constraints from XENON100 are, as a result of the factor $v_{\rm rel}^2$ in \eqref{eq:sigmaSIbounds}, comparable to the  Fermi LAT constraints if $m_\chi \lesssim 40 \, {\rm  GeV}$. For light DM, direct detection experiment hence start to indirectly probe annihilation cross sections of almost $\langle  \sigma v_{\rm rel} \rangle_{\gamma \gamma} \,  \sim  \, 10^{-29}  \,  {\rm cm}^3 \, {\rm s}^{-1}$. This feature is illustrated  by the red solid and the black dashed line in the right panel of Figure~\ref{fig:bounds}. Notice that towards larger masses, the bound becomes weaker in spite of the stronger limit on $\mathcal{C}^\chi$, because of the factor of $m_\chi^4$ appearing in \eqref{eq:sigmaSIbounds}.

Nevertheless, if DM is fermionic and annihilates into $\gamma$-rays via the operator $\mathcal{O}^\chi$, the tentative $\gamma$-ray signal at $m_\chi = 130$ GeV will be easily within reach of XENON1T~\cite{Aprile:2012zx} after a run-time of one year. This is indicated by the purple dotted line in the right panel of Figure~\ref{fig:bounds}. For this estimate, we have assumed that the XENON1T detector has identical properties as XENON100, except for the fiducial mass, which we take to be 1.1 tons. Assuming different detector properties close to the threshold does not affect the bounds for $m_\chi \gtrsim 100$ GeV. On the other hand, since we expect a significant number of events at large recoil energies, raising the upper bound of the DM search window to larger values of  the primary  scintillation signal $S1$ would additionally increase the sensitivity of XENON1T. We have checked that the uncertainties in  $v_\text{rel}$, $f^N_{Tq}$ and $\mathcal{L}_\text{eff}$ do not compromise our results.

\section{Discussions}
\label{sec:discussion}

Motivated by recent results from searches for $\gamma$-ray lines and tentative claims of a signal, we have studied the direct detection cross section arising from interactions of DM with two photons. The di-photon annihilation cross section can be cast in terms of effective operators which allows for a translation into direct detection event rates. To this end we identified all effective operators giving rise to DM annihilation into two photons up to dimension 6 for real and complex scalar DM and up to dimension 7 for Dirac and Majorana DM and calculated the loop-induced scattering cross section in a model-independent way.  Our operator analysis is valid under the assumption that the scale of new physics is sufficiently high and in particular that DM does not annihilate via an $s$-channel resonance.

The loop-induced direct detection cross section  receives contributions from two different types of effects. The first  accounts for the fact that loop diagrams involving  high-virtuality photons induce couplings of DM to quarks and gluons, while the second one corresponds to Rayleigh scattering  that describes the low-energy interactions of the photons with the total electric charge of the nucleus. We discussed in detail the QED and QCD mixing  of operators that leads to the former corrections, which are typically neglected in the direct detection literature. While the resulting contributions have the standard $A^2$ scaling, the Rayleigh contribution scales as $Z^4$.

At zero-momentum transfer the nuclear matrix elements for direct detection on xenon are comparable in size but have opposite signs, which leads to destructive interference.  For finite momentum transfer, however, the form factors of the two contributions behave rather differently~-- the Rayleigh form factor gets suppressed much faster such that the overall event rates in xenon are dominated by the contribution due to mixing. Furthermore, the resulting recoil spectrum has a dip just at the lower end of the search window~-- a striking feature that, if observed, could confirm the interference of the two different form factors. For lighter targets Rayleigh scattering is more strongly suppressed due to the $Z^4$ dependence,  while on the other hand the nuclear coherence scale $Q_0$ is larger which eases the form factor suppression for finite momentum transfer.

The overall scale of the DM-nucleus cross section is proportional to the annihilation cross section into two photons and also depends on the leading partial wave of the annihilation process. For  $\langle \sigma_{\chi} \hspace{0.25mm} v_{\text{rel}} \rangle_{\gamma \gamma} \, \sim \, 10^{-27} \, \text{cm}^3 \, \text{s}^{-1}$, the induced direct detection cross section for  $s$-wave annihilation is too small to be observed in upcoming direct searches, even for spin-independent cross sections. Operators which lead to spin- or momentum-dependent cross sections are even less constrained by direct detection experiments. For $p$-wave annihilation, on the other hand, the DM nucleus cross section is just below the sensitivity of XENON100, but within reach of XENON1T. 

\section*{Acknowledgements}

We thank Vincenzo Cirigliano and Neal Weiner for useful discussions. KSH acknowledges support from ERC Advanced Grant BSMOXFORD 228169. FK is supported by the Studienstiftung des Deutschen Volkes. We thank the European Research and Training Network ``Unification in the LHC era'' (PITN-GA-2009-237920) for partial support. 

\appendix

\section{Velocity dispersion of Galactic DM}
\label{app:vrel}
Given a mass model of the Galaxy, consisting of a DM distribution $\rho_\chi(r)$ as well as baryonic contributions, the dispersion $\sigma_{\chi,r}^2 =\langle v_{\chi,r}^2 \rangle$ of the DM radial velocity $v_{\chi,r}$ follows from the Jeans equation. In particular, assuming spherical symmetry and isotropy of the distribution function~\cite{Binney:2008zz},
\begin{equation}
\sigma_{\chi,r}^2(r) = \frac{1}{\rho_\chi(r)} \int_r^{\infty} d r' \rho_\chi(r') \, \frac{d \Phi (r')}{{d}r'} \, ,
\end{equation}
where $\Phi(r)$ is the gravitational potential generated by all mass components. The speed $v_{\chi} = \sqrt{v_{\chi,r}^2 + v_{\chi,\phi}^2 + v_{\chi,\theta}^2}$ is distributed as a Maxwell-Boltzmann distribution proportional to $v_{\chi}^2 \exp{ [ - v_{\chi}^2 / v_0^2 ] }$ with $\langle v_{\chi}^2 \rangle = 3 v_0^2 /2$ and  $v_0= 2 \sigma_{\chi,r}(r)$. The dispersion of the relative velocity $v_{\text{rel}} = |\vec{v}_{\chi,1} - \vec{v}_{\chi,2}|$ is simply $\langle v_{\text{rel}}^2 \rangle = \langle (\vec{v}_{\chi,1})^2 \rangle + \langle (\vec{v}_{\chi,2})^2 \rangle - 2 \langle \vec{v}_{\chi,1} \cdot \vec{v}_{\chi,2} \rangle = 2 \langle v_{\chi}^2 \rangle = 3 v_0^2$, where we have used that $\langle \vec{v}_{\chi,1} \cdot \vec{v}_{\chi,2} \rangle$ must vanish by symmetry. For our analysis we adopt the benchmark value of $v_0 = 220 \, \text{km} \, \text{s}^{-1}$, which gives $v_\text{rel} = \sqrt{3} v_0 \simeq 1.3 \cdot 10^{-3} \, c$.

In general, however, the velocity distribution of DM is a function of the galacto-centric radius $r$. For $s$-wave annihilation, one has $\langle \sigma_\chi v_{\mathrm {rel}} \rangle = \mathrm{const.}$ such that the $r$-dependence does not affect our results. For $p$-wave annihilation, however, one needs to estimate by how much the $r$-dependence of the velocity distribution enhances or suppresses the velocity-averaged annihilation cross section.

To this end, we employed two state-of-the-art models~\cite{Iocco:2011jz} (model 1 and model 5) for the baryonic mass distribution of the Galaxy which have been shown to fit microlensing as well as rotation curve measurements. Model 1 consists of a bulge, a thin and a thick disk while model 5 consists of a bulge, a bar, a stellar disk as well as a gas disk. For both baryonic models we considered a Navarro-Frenk-White (NFW) profile~\cite{Navarro:1995iw} $\rho_\chi \propto x^{-\alpha} (1+x)^{-3+\alpha}$ (where $x=r/r_s$, $r_s = 20 \, \text{kpc}$ and $\alpha=1$) as well as an Einasto profile~\cite{Graham:2005xx} $\rho_\chi \propto \exp{ [-2 (x^\alpha-1) / \alpha ] }$ (where $r_s$ as above and $\alpha = 0.17$) with the local DM density fixed to $\rho_\text{DM} = 0.4 \, \text{GeV} \, \text{cm}^{-3}$. We compared $\langle \sigma_\chi v_{\rm rel} \rangle \rho_\chi^2$ for the case of the full $r$-dependence to the case where only $\rho_\chi^2$ depends on $r$ and $\langle \sigma_\chi v_{\rm rel} \rangle$ is fixed to its local value. In the inner kiloparsec, the velocity dependence leads to a suppression proportional to $r^{0.5 \mathellipsis 0.85}$, but when integrating over lines-of-sight and averaging over the target regions of~\cite{Weniger:2012tx} the contribution from the inner kiloparsecs is strongly suppressed. We find an overall enhancement of the velocity-averaged annihilation cross section with respect to the case with a fixed velocity distribution for an NFW (Einasto) profile of $15 \%$ ($5 \%$) for model 1 and $27 \%$ ($30 \%$) for model 5. This enhancement can be traced back to the pronounced bulge/bar, especially in model 5, but barring the uncertainties of the mass modelling in the inner few kiloparsecs of the Galaxy, we chose to ignore the uncertainty related to this effect in our final results.

\section{Mixing and matching of loop-induced DM operators}
\label{app:running}

In this appendix, we review in detail the mixing and matching of the following effective DM operators, 
\begin{equation} \label{eq:app1}
{\cal O}^\chi = {\mathcal C}^\chi \chi \chi F^{\mu \nu} F_{\mu \nu} \,, \qquad
{\cal O}_{q}^\chi = {\mathcal C}_{q}^\chi  m_q \chi \chi \bar q q \,, \qquad 
{\cal O}_{G}^\chi = {\mathcal C}_{G}^\chi  \chi \chi G^{a, \mu \nu} G_{\mu \nu}^a \,. 
\end{equation}
Since by assumption the DM particle $\chi$ is a singlet under the SM gauge group, the $\chi \chi$ parts of the above composite operators can be ignored for the further discussion. This feature implies that the RG evolution and the matching corrections to the Wilson coefficients in~\refeq{eq:app1} resemble those well-known from Higgs physics. 

In fact, the renormalisation scale dependence of the Wilson coefficients  is fully determined in terms of the QED  beta function $\beta_{e}$ and the corresponding mass anomalous dimensions $\gamma_m^e$ of the quarks
\begin{equation} \label{eq:app2}
\begin{split}
& \frac{d }{d \ln \mu} \, \alpha(\mu)  =  \beta_e (\mu) = -2 \alpha (\mu) \sum_{n=0}^\infty \beta_e^{(n)} \left ( \frac{\alpha (\mu)}{4 \pi} \right )^{n+1} \,, \\[1mm]
& \hspace{1mm} \frac{d}{d\ln \mu} \, \ln m (\mu) =  \gamma_m^e (\mu) = -\sum_{n=0}^\infty \gamma_{m}^{e (n)} \left ( \frac{\alpha (\mu)}{4 \pi} \right )^{n+1} \,,
\end{split}
\end{equation}
as well as their QCD counterparts $\beta_s$ and $\gamma_m^s$. Note that for our definition of the beta functions, $\beta_e (\mu)  > 0$ while $\beta_s (\mu) < 0$. Assuming that the Wilson coefficient ${\cal C}_G^\chi$ vanishes at the new-physics scale $M_\ast$, one obtains for scales $m_t < \mu < M_\ast$ the expressions
\begin{equation} \label{eq:app3}
\begin{split}
{\cal C}^\chi (\mu ) & = \frac{\alpha (M_\ast) \hspace{0.25mm} \beta_e (\mu)}{\alpha (\mu) \hspace{0,25mm} \beta_e (M_\ast)} \, {\cal C}^\chi (M_\ast ) \,, \\[1mm]
{\cal C}_q^\chi (\mu)  & = \frac{4 \alpha (M_\ast)}{\beta_e (M_\ast)} \, \big (  \gamma_m^e (M_\ast) - \gamma_m^e (\mu) \big ) \, {\cal C}^\chi (M_\ast) + {\cal C}^\chi_{q} (M_\ast)\,.
\end{split}
\end{equation}
These all-order results represent the QED analogs of the formulas given in~\cite{Chetyrkin:1996ke} for the case of QCD. Here $\alpha$, $\beta_e$, and $\gamma_m^e$ all correspond to quantities with the number of flavours set to those active at  $\mu$. Expanding \refeq{eq:app3} in powers of the electromagnetic coupling $\alpha$ and keeping only terms relevant for our analysis, one finds 
\begin{equation} \label{eq:app4}
\begin{split}
{\cal C}^\chi (\mu ) & \simeq \frac{\alpha (\mu)}{\alpha (M_\ast)} \, {\cal C}^\chi (M_\ast ) \,, \\[1mm]
{\cal C}_q^\chi (\mu)  & \simeq \frac{2 \gamma_m^{e (0)}}{\beta_e^{(0)}} \left ( 1 -  \frac{\alpha (\mu)}{\alpha (M_\ast)} \right ) {\cal C}^\chi (M_\ast) \,, 
\end{split}
\end{equation}
for the case ${\cal C}^\chi_{q} (M_\ast) = 0$. Notice that the appearance of the rescaling factor $\alpha (\mu)/\alpha (M_\ast)$ in the first expression  guarantees the proper normalisation of the matrix element of ${\cal O}^\chi$ at the  scale $\mu$. Using now the LO relations
\begin{equation} \label{eq:app5}
\frac{\alpha (\mu)}{\alpha (M_\ast)} \simeq 1  + \frac{\alpha (M_\ast)}{4 \pi} \, \beta_e^{(0)} \, \ln \left ( \frac{M_\ast^2}{\mu^2} \right ) \,, 
\end{equation}
and $\gamma_m^{e (0)} = 6 e_q^2$ the results in \refeq{eq:app4} can be further simplified. In particular, neglecting small effects associated to the running of $\alpha$, one arrives at \refeq{eq:CqM}. 

At each heavy-quark threshold $m_Q$ the Wilson coefficient ${\cal C}_G^\chi$ receives finite matching corrections that are known to NNNLO \cite{Schroder:2005hy,Chetyrkin:2005ia}. In the case of $m_t$, one has including NNLO contributions \cite{Kramer:1996iq,Chetyrkin:1997iv}
\begin{equation} \label{eq:app6}
{\cal C}_G^\chi (m_t) = -\frac{\alpha_s (m_t)}{12 \pi} \left ( 1 + \frac{11\alpha_s (m_t)}{4 \pi} + \frac{\alpha_s^2 (m_t) \left ( 2777 - 201 n_f \right ) }{288 \pi^2} \right )  {\cal C}_t^\chi (m_t) \,. 
\end{equation} 
The same expression applies to the bottom- and charm-quark thresholds $m_b$ and $m_c$ after adjusting the number of active flavours  $n_f$. Since the second, flavour-dependent term in brackets represents only a relative correction of below $1\%$, we neglect such terms in our numerical analysis. Likewise, we do not include small finite threshold corrections to the Wilson coefficients ${\cal C}_q^{\chi}$ that start at order $\alpha_s^2$~(see for example \cite{Chetyrkin:1996ke}). 

Once the top quark has been integrated out, one has to consider the mixing of the full set of operators \refeq{eq:app1}. The solutions to the RG equations read in this case 
\begin{equation} \label{eq:app7}
\begin{split}
{\cal C}^\chi (\mu ) & = \frac{\alpha (m_t) \hspace{0.25mm} \beta_e (\mu)}{\alpha (\mu) \hspace{0,25mm} \beta_e (m_t)} \, {\cal C}^\chi (m_t) \,, \\[1mm]
{\cal C}_q^\chi (\mu)  & = \frac{4 \alpha (m_t)}{\beta_e (m_t)} \, \big ( \gamma_m^e (m_t) - \gamma_m^e (\mu) \big ) \, {\cal C}^\chi (m_t) \\ & \phantom{xx} + \frac{4 \alpha_s (m_t)}{\beta_s (m_t)} \, \big ( \gamma_m^s (m_t) - \gamma_m^s (\mu) \big ) \, {\cal C}^\chi_{G} (m_t) + {\cal C}^\chi_{q} (m_t)\,, \\[1mm]
{\cal C}_{G}^\chi (\mu ) & = \frac{\alpha_s (m_t) \hspace{0.25mm} \beta_s (\mu)}{\alpha_s (\mu) \hspace{0,25mm} \beta_s (m_t)} \, {\cal C}_{G}^\chi (m_t) \,, \\[1mm]
\end{split}
\end{equation}
where $m_b < \mu < m_t$. Performing again a series expansion in the coupling constants, using the QCD version of \refeq{eq:app5} and $\gamma_m^{s (0)} = 8$, one derives from \refeq{eq:app7} the approximations
\begin{eqnarray} \label{eq:app8}
\begin{split}
{\cal C}^\chi (\mu) & \simeq \frac{\alpha (\mu)}{\alpha (M_\ast)} \, {\cal C}^\chi (M_\ast ) \,, \\[1mm]
{\cal C}_q^\chi (\mu) & \simeq \left [ -3  e_q^2 \hspace{0.5mm} \frac{\alpha (M_\ast)}{\pi}  \ln \left ( \frac{M_\ast^2}{\mu^2} \right ) - e_t^2   \hspace{0.5mm}  \frac{\alpha (M_\ast)}{\pi} \! \left (  \frac{\alpha_s (m_t)}{\pi} \right )^2   \ln \left ( \frac{m_t^2}{\mu^2} \right )   \ln \left ( \frac{M_\ast^2}{m_t^2} \right ) \right ] \hspace{-0.25mm} {\cal C}^\chi (M_\ast) \,, \hspace{7mm} \\[1mm]
{\cal C}_{G}^\chi (\mu) & \simeq  \frac{\alpha (M_\ast)}{\pi} \, \frac{\alpha_s (\mu)}{\pi} \, \frac{e_t^2}{4} \left (1 + \frac{11\alpha_s(m_t)}{4\pi} \right )  \ln \left ( \frac{M_\ast^2}{\mu^2}\right )  {\cal C}^\chi (M_\ast) \,.  
\end{split}
\end{eqnarray}
The appearance of the overall factor $\alpha_s (\mu)$ in ${\cal C}_G^\chi$ is again crucial for the correct normalisation of the matrix element of ${\cal O}_G^\chi$. Notice that the second term in the square bracket  entering the expression for ${\cal C}_q^\chi$ is subleading and can be neglected for all practical purposes.  Following the above discussion the implementation of the bottom- and charm-quark thresholds is straightforward. Neglecting subleading  $\mathcal{O}(\alpha_s^2)$ terms in \refeq{eq:app8} as well as the evolution of $\alpha$, which is a very good approximation, the final expressions for the low-energy Wilson coefficients given in~\refeq{eq:Clow} are then readily obtained. 

\section{Direct detection for XENON}
\label{app:xenon}

The typical recoil energy in a given direct detection experiment is determined both by the expected recoil spectrum for DM interactions with the detector material and the properties of the detector itself. The recoil spectrum is proportional to the velocity integral
\begin{equation} \label{eq:gvt} 
g (v_\text{min}) = \int_{v_\text{min}}^\infty \! d^3 v \; 
\frac{f(\boldsymbol{v} + \boldsymbol{v}_\mathrm{E})}{v}\, ,
\end{equation}
where $f(v)$ is the local DM velocity distribution evaluated in the Galactic rest frame, $\boldsymbol{v}_\mathrm{E}$ is the velocity of the Earth relative to this frame, and $v=|\boldsymbol{v}|$.  The minimum velocity required for a DM particle to transfer an energy $E_\text{R}$ to a nucleus is  given by $v_\text{min}(E_\text{R}) = \sqrt{m_A E_\text{R} / (2\mu_A^2)}$.

The performance of the detector, on the other hand, can be characterised by the response function $\text{Res}\left (E_\text{R} \right )$. This function encodes the probability that a scattering process with recoil energy $E_\text{R}$ will lead to an observable signal within the DM search window of the experiment. For the XENON100 experiment, the search window is defined in terms of the primary scintillation signal $S1$ by the condition that it should consist of between 3 and 20 photoelectrons (pe)~\cite{Aprile:2012}.

\begin{figure}
\begin{center}
\includegraphics[width=0.45\columnwidth,clip,trim=0 -14 0 0]{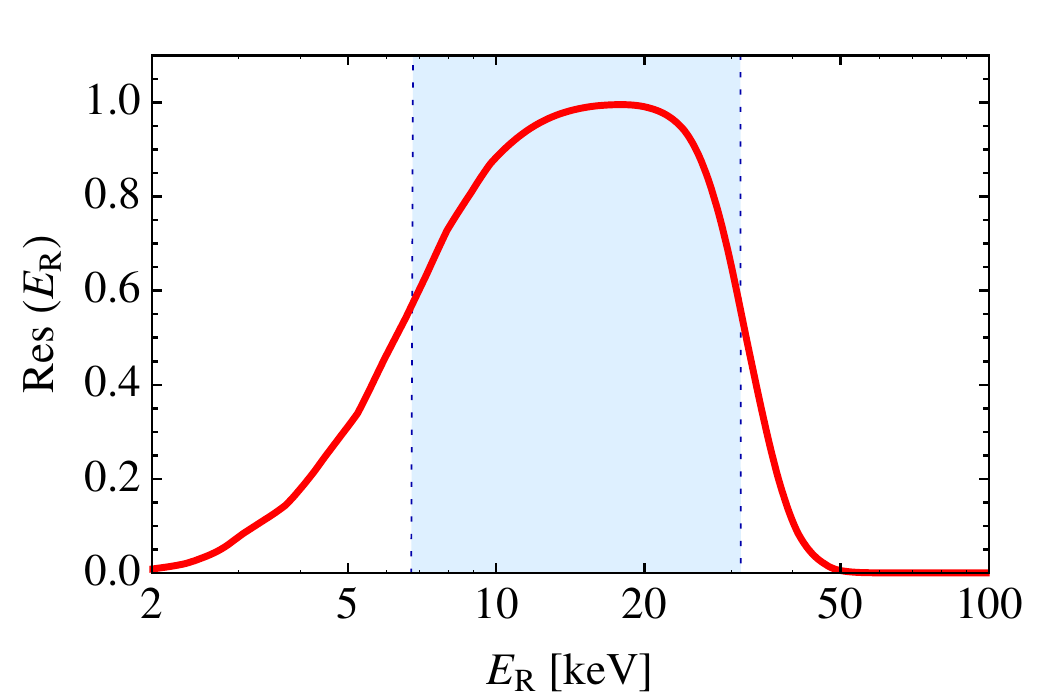}
\quad 
\includegraphics[width=0.5\columnwidth]{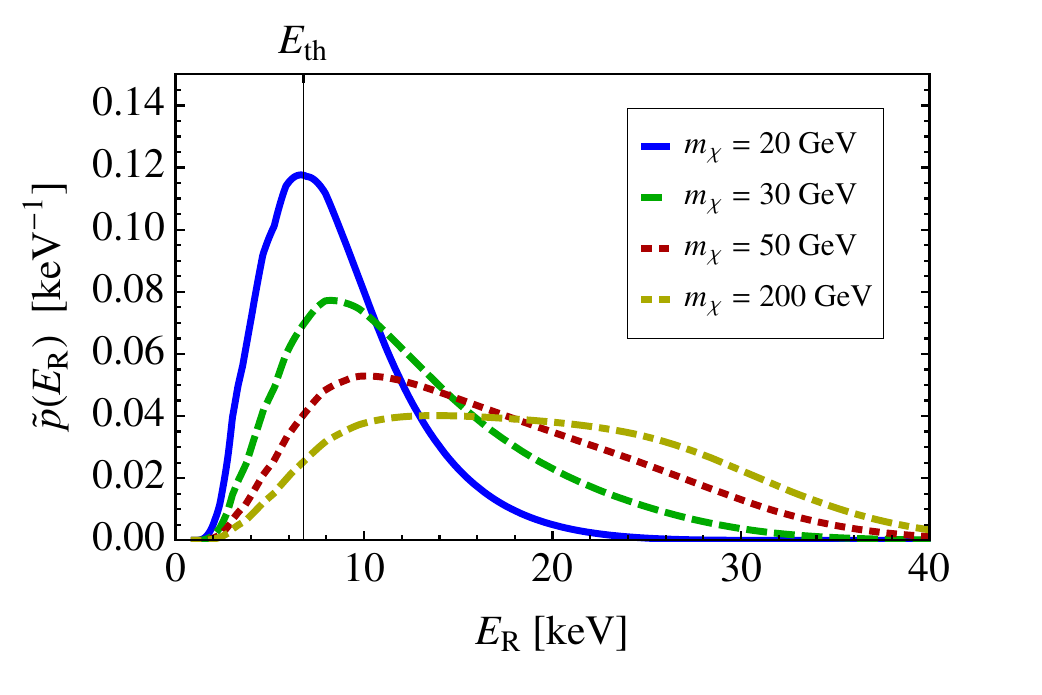}
\caption{\label{fig:1} Left panel: Response function $\text{Res} \left (E_\text{R} \right )$ for XENON100. The blue shaded region displays the acceptance window in the absence of fluctuations. Right panel:  Normalised probability distribution of recoil energies $\tilde{p}(E_\text{R})$ for different values of the DM mass.}
\end{center}
\end{figure}

For a given recoil energy, the expected number of $S1$ photoelectrons is 
\begin{equation} \label{eq:S1ER}
S1(E_\text{R}) = E_\text{R} L_y \mathcal{L}_\text{eff} \, \frac{S_\text{nr}}{S_\text{ee}} \, ,
\end{equation}
where $L_y = 2.28 \, {\rm pe} \, {\rm keV}^{-1}$ is the light yield of the detector, $\mathcal{L}_\text{eff}$ is the relative scintillation efficiency of liquid xenon,  while $S_\text{nr} = 0.95$ and $S_\text{ee} = 0.58$ are the field-dependent quenching factors for nuclear and electron recoils, respectively. These values are taken from~\cite{Aprile:2012}, while the best-fit curve for $\mathcal{L}_\text{eff}$ is taken from Figure~1 of~\cite{Aprile:2011hi}.
Due to fluctuations, the actually observed number of photoelectrons in a collision may deviate significantly from the expectation value~\cite{Sorensen:2010hq}. Assuming that the distribution of photoelectrons can be described with Poisson statistics, the probability to observe $n$ photoelectrons in a collision with recoil energy $E_\text{R}$ is hence given by
\begin{equation} \label{eq:fnER}
f(n,E_\text{R}) = \frac{\exp \left (-S1(E_\text{R}) \right ) S1(E_\text{R})^n}{n!} \, .
\end{equation}
Consequently, the probability of a nuclear recoil with energy $E_\text{R}$ to give a primary scintillation signal  $S1$ within the search window is 
\begin{equation}
\text{Res} \left (E_\text{R} \right ) = \sum_{n=3}^{20} f(n,E_\text{R}) \, .
\end{equation}
The response function for XENON100 is shown in the left panel of Figure~\ref{fig:1}. For comparison also the acceptance window in the absence of fluctuations  is indicated.

It follows, that the observed distribution of recoil energies in the XENON100 detector is then given by the product
\begin{equation} \label{eq:pER} 
p(E_\text{R}) = g (v_\text{min}(E_\text{R})) \, \text{Res} \left (E_\text{R} \right ) \,,
\end{equation}
where  the energy dependence of the cut acceptance in XENON100 has been neglected for simplicity. For light DM, $g (v_\text{min}(E_\text{R}))$ is a steeply falling function of $E_\text{R}$, whereas the response function $\text{Res} \left (E_\text{R} \right )$ is strongly suppressed for $E_\text{R} < E_\text{th}$ with the threshold energy  determined by the requirement that $S1(E_\text{th})=3$ pe. Consequently,  the function $p(E_\text{R})$ will be strongly peaked at  $E_\text{R} \simeq E_\text{th}$  and as a result the typical momentum transfer for light  DM amounts to $q \simeq \sqrt{2 m_A E_\text{th}}$. 

For heavy DM, the function $g(v_\text{min} (E_{\rm R}))$ decreases much more slowly and one needs a slightly more elaborate treatment. In order to deal with this case, we first define the normalised distribution of recoil energies
\begin{equation} \label{eq:ptilde} 
\tilde{p}(E_\text{R}) = \frac{p(E_\text{R})}{\int_0^\infty d E_\text{R} \, p(E_\text{R})} \,.
\end{equation}
Since $v_\text{min} (E_{\rm R})$ depends indirectly on the mass $m_\chi$ of the DM particle, so does $\tilde{p}(E_\text{R})$. The function $\tilde{p}(E_\text{R})$ is shown in the right panel of Figure~\ref{fig:1} for different values of $m_{\chi}$. One observes that for increasing DM mass the peak of the distribution is not only shifted to higher recoil energies, but that also the tail of the spectrum becomes much more pronounced. For an accurate description, one hence has to use  momentum-averaged form factors
\begin{equation} \label{eq:averagedFF} 
\langle F \rangle = \int_0^\infty d E_\text{R} \, \tilde{p}(E_\text{R}) \hspace{0.25mm}  F (E_\text{R})  \, .
\end{equation}

\providecommand{\bysame}{\leavevmode\hbox to3em{\hrulefill}\thinspace}

\end{document}